\documentclass{jfm}
\usepackage[pdftex]{graphicx,color}
\usepackage{amsmath}
\usepackage{amssymb}
\usepackage{color}
\usepackage{epsfig}
\usepackage{natbib}
\usepackage{eepic}
\usepackage{vmargin}
\ifCUPmtlplainloaded \else
 \checkfont{eurm10}
 \iffontfound
 \IfFileExists{upmath.sty}
  {\typeout{^^JFound AMS Euler Roman fonts on the system,
     using the 'upmath' package.^^J}%
  \usepackage{upmath}}
  {\typeout{^^JFound AMS Euler Roman fonts on the system, but you
     dont seem to have the}%
  \typeout{'upmath' package installed. JFM.cls can take advantage
     of these fonts,^^Jif you use 'upmath' package.^^J}%
  }
 \else
 \fi
\fi

\ifCUPmtlplainloaded \else
 \checkfont{msam10}
 \iffontfound
 \IfFileExists{amssymb.sty}
  {\typeout{^^JFound AMS Symbol fonts on the system, using the
    'amssymb' package.^^J}%
  \usepackage{amssymb}%

  }{}
 \fi
\fi

\ifCUPmtlplainloaded \else
 \IfFileExists{amsbsy.sty}
 {\typeout{^^JFound the 'amsbsy' package on the system, using it.^^J}%
  \usepackage{amsbsy}}
 {}
\fi

\newcommand\Imag{\mbox{Im}} 

\newsavebox{\astrutbox}
\sbox{\astrutbox}{\rule[-5pt]{0pt}{20pt}}

\newcommand\tti{\ensuremath{\rightarrow\infty}}

\newcommand\ttz{\ensuremath{\rightarrow 0}}

\title[]
{The growth of wind-waves in finite depth}
 \author[P. Montalvo, J. Dorignac, M. A. Manna, C. Kharif and H. Branger]
 {P. \ls \ls M\ls O\ls N\ls T\ls A\ls L\ls V\ls O$^{1,2}$, \ns
J.\ns \ls D\ls O\ls R\ls I\ls G\ls N\ls A\ls C\ls $^{1,2}$,
 \ns M.\ls \ns A.\ns M\ls A\ls N\ls N\ls A$^{1,2}$ 
 \thanks{Email address for correspondence: Miguel.Manna@univ-montp2.fr} 
 , \ns C.\ns K\ls H\ls A\ls R\ls I\ls F\ls $^{3,4}$ 
and H.\ns BRANGER\ls$^4$}
 \affiliation{$^1$
 Universit\'e Montpellier 2, Laboratoire Charles Coulomb UMR 5221, F-34095, Montpellier, France. \\[\affilskip]
$^2$ CNRS, Laboratoire Charles Coulomb UMR 5221, F-34095, Montpellier, France.\\
$^3$ Ecole Centrale Marseille, 38 rue Fr\'ed\'eric Joliot-Curie, 13451 Marseille cedex 20.\\
 $^4$Institut de Recherche sur les Ph\'enom\`enes Hors \'Equilibre, CNRS, AMU, UMR 7342,
 49 rue Fr\'ed\'eric Joliot-Curie, BP 146, 13384 Marseille cedex 13, France.
 \\[\affilskip]}
 \date{June 2012}

\begin{document}
\maketitle
\begin{abstract}
In order to study the growth of wind waves in finite depth we extend Miles' theory to the finite depth domain. A depth-dependent wave growth rate is derived from the dispersion relation of the wind/water interface. A suitable dimensionless finite depth wave age parameter allows us to plot a family of wave growth curves,
each family member characterized by the water depth. Two major results are that
for small wave age, the wave growth rates are comparable to those of deep water and
for large wave age, a finite-depth wave-age-limited growth is reached, with wave growth rates going to zero. The corresponding limiting wave length and limiting phase speed are explicitely calculated in the shallow and in the deep water cases.
A qualitative agreement with well-known empirical results is established and shows the robust consistency of the linear theoretical approach.
\end{abstract}
\section{Introduction}
The problem of wind-induced wave growth, whose starting points are the
Navier-Stokes equations, is a formidable one and several approximations and
assumptions were done to fully understand it. The pioneer works are those of 
\citet{Jeffreys1, Jeffreys2}, Phillips (1957) and \citet{miles2}. 

From a mathematical point of view, Miles' theory is based on the dispersion relation of the air-sea interface and
the related Rayleigh equation \citep{Rayleigh,ConteMiles,DrazinReid}.
Miles' physical mechanism of wave generation by wind assumes that ocean surface waves
are generated by a resonance phenomenon. Resonance appears between the
wave-induced pressure gradient on the inviscid shear airflow and the surface waves. Basically, it
occurs when the phase velocity of the surface wave equals the flow velocity.
The usual model has two spatial dimensions, the water is assumed deep, the viscosity is disregarded, 
the equations of motions are linearized and  air flow turbulence is largely neglected, except in its role in setting up the logarithmic shear flow.
The air domain is also considered inviscid and the dynamics 
equations are linearized around a prescribed mean wind velocity. The subsequent Rayleigh equation, depending on a logarithmic wind profile, is solved by a combination of analytical and numerical methods. 
For a good review on wind-induced wave growth as well as further progress in quasi linear 
theory see \citet{Miles3} and \citet{Janssen}. Alternatively, a review of Jeffreys' theory and some of its applications can be found in \citet{TouboulKharif1}.

Despite these severe simplifications, the Miles' theory brings a sufficiently ideal model to investigate the physics of the water wave growth problem. It allows, at least linearly, an analytical treatement of the phenomenon.

Nonetheless, Miles' theory is limited to deep water thus restrictive with regards to winds generating nearshore ocean waves or shallow lakes waves. This issue is really challenging the physics community and the engineering
 community. The former because the only way to tackle this basic question in interfacial fluid dynamics is
  \emph{via} an adequate modelling, and the latter because in coastal engineering works the wave field must 
  be influenced by the water depth. Therefore a theoretical extension of Miles's mechanism to finite depth is lacking in the field.

The situation is very different from the experimental point of view. A lot of investigations were done on the subject. Especially the experiments in the shallow Lake George in Australia by Young and co-workers in references \citet{Young1}, \citet{Young2}, 
\citet{YoungVerhagen1} and \citet{YoungVerhagen2}. They gave one of the first systematic attempts to understand the physics of wave-wind generation in water of finite depth. They provided empirical relationships in terms of appropriate dimensionless parameters able to reproduce experimental data. Fetch limited wind-waves growth experiments in finite depth water
confirmed the existence of asymptotic limits to growth of wave energy and spectrum peak frequency. The more important discovery
was that at short fetches the growth rates tended to be comparable to the deep water limit. At longer fetches, the growth rate is lower in shallow water than in deep water, and beyond a limit fetch, the growth rate vanishes.

In this paper we deal with the extension of Miles' theory to the finite depth context. 
We  use the Rayleigh equation associated to the air motion with the linear dispersion relation adapting the method given by \citet{BejiN} for the deep water case. An suitable non-dimensional finite depth wave age allow us to derive an explicit expression for the wave growth rate. Evolution of the latter against the wave age is plotted. It clearly shows a phenomenon of finite depth limited growth. The model is able to derive from theoretical expressions (with a good degree of accuracy) important empirical results such as the asymptotic limit where wave growth rate goes to zero.

The paper is organized as follows. 
In sections  (\ref{intro}) and (\ref{water domain}) the linear problem in the water domain is introduced after a brief presentation of the model. In section (\ref{air domain})
the air domain is coupled with the water domain and make it possible to solve the linear interface problem.
We derive the linear dispersion relation of the Miles' theory of wind wave generation in finite depth. In section (\ref{growth}) we introduce dimensionless variables and scalings and we obtain the suitable growth rate. The linear dynamic is discussed in section (\ref{Results}) with special emphasis on its qualitative agreement
with empirical relations. Section (\ref{conclusions}) draws the conlusions and the perspectives. Finally in Appendix (\ref{Orr}) we give some details of the numerical work behind.
\section{The interface problem.}
\label{intro}
We are going to study the stability of an air-water interface. 
Let the fluid particles be located relatively to a fix rectangular 
Cartesian frame with origin $O$ and axes $(x, y, z)$, where 
$Oz$ is the upward vertical direction. We assume translational symmetry along $y$ 
and we will only consider a sheet of fluid parallel to the $xz$ plane. 
$z=0$ characterizes the interface at rest. The perturbed air-water interface will be described by $z=\eta(x,t)$.
The air occupies the $\eta(x,t)<z<+\infty$ region, and the water lies between 
the bottom located at $z=-h$ and the free surface $z=\eta(x,t)$. We assume the water as well as the air inviscid and incompressible. The air flow will be described by a prescribed mean shear flow, only depending on the vertical
coordinate $z$. We assume the dynamic to be linear, and disregard the air turbulence, building a \textit{quasi-laminar theory}.

\subsection{The water domain}
\label{water domain}
In the water domaine we consider the Euler equations for finite depth.
The horizontal and vertical velocities of the fluid are $u(x,z,t)$, and $w(x,z,t)$.
The continuity equation and the linearized equation of motion in the water domain read as (\citet{Lighthill})
\begin{equation}\label{EulerWater}
u_x + w_z = 0, \quad \rho_w u_t= -P_{x},\quad \rho_w w_t=-P_{z}-g \rho_w,
\end{equation}
where $P(x,z,t)$ is the pressure, $g$ the gravitational acceleration, $\rho_w$ is the water density and
subscripts in $u$, $w$ and $P$ denote partial derivatives.
The boundary conditions at $z = -h $ and at $z = \eta (x,t)$ are
\begin{eqnarray}
w(-h)&=& 0,\quad \eta_{t}=w(0), \label{EulerWaterBC1}\\
P(x,\eta,t)&=& P_a(x,\eta,t),\label{EulerWaterBC2}
\end{eqnarray}
where $P_a$ is the air pressure evaluated at $z=\eta$. Thus equation (\ref{EulerWaterBC2}) is the continuity of the pressure across the  air/water interface. As this is a vital assumption for the growth mechanism, we give it a more pleasant expression. So, let us introduce a reduced pressure defined by
\begin{equation}\label{reduced pressure}
\mathbf{P}(x,z,t)=P(x,z,t) + \rho_w gz -P_{0},
\end{equation}
where $P_0$ is the atmospheric pressure. In terms of (\ref{reduced pressure}) 
equations (\ref{EulerWaterBC1})-(\ref{EulerWaterBC2}) read
\begin{equation}\label{EulerWaterReduced}
u_x + w_z = 0, \quad \rho_w u_t= -\mathbf{P}_x,\quad \rho_w w_t=-\mathbf{P}_z,
\end{equation}
\begin{eqnarray}
w(-h)&=& 0,\quad \eta_{t} = w(0), \label{EulerWaterReducedBC1} \\
\mathbf{P}(x,\eta,t)&=&P_a(x,\eta,t)+\rho_w g\eta - P_0. \label{EulerWaterReducedBC2}
\end{eqnarray}
The linear equations system (\ref{EulerWaterReduced})-
(\ref{EulerWaterReducedBC2}) can be solved, assuming normal mode solutions 
as
\begin{equation}\label{Fourier}
 \mathbf{P}=\mathcal{P}(z)\exp{(i\theta)},\quad u=\mathcal{U}(z)\exp{(i\theta)},
 \quad w=\mathcal{W}(z)\exp{(i\theta)},\quad
\eta=\eta_{0}\exp{(i\theta)},
\end{equation}
with $\theta= k(x-ct)$ where $k$ is the wavenumber, $c$ the phase speed and $\eta_0$ is a constant.
Using equations (\ref{EulerWaterReduced}), (\ref{EulerWaterReducedBC1}), 
(\ref{EulerWaterReducedBC2}) and (\ref{Fourier}) we obtain  
\begin{eqnarray}
w(x,z,t) &=&\frac{-ikc\sinh{k(z+h)}}{\sinh{kh}}\eta_0\exp{(i\theta)},\label{wlinear}\\
u(x,z,t) &=& \frac{kc\cosh{k(z+h)}}{\sinh{kh}}\eta_0\exp{(i\theta)}, \label{ulinear}\\
\mathbf{P}(x,z,t) &=& \frac{k\rho_wc^2\cosh{k(z+h)}}{\sinh{kh}}\eta_0\exp{(i\theta)}\label{Plinear}.
\end{eqnarray}
The phase speed $c$ is unknown in equations (\ref{wlinear})-(\ref{Plinear}). 
To determine $c$ we have to consider the boundary conditions (\ref{EulerWaterReducedBC2}) (not yet used) and (\ref{EulerWaterReducedBC1}) which yields
\begin{equation}\label{deterc}
 c^2k\rho_w\eta_0\exp{(i\theta)}\coth{kh}-\rho_w g\eta_0\exp{(i\theta)} + P_0=P_a(x,\eta,t).
\end{equation}
In the single-domain problem $P_a(x,\eta,t)= P_0$ and (\ref{deterc}) gives the usual expression for c,
\begin{equation}
 c^2 = c_0^2 =\frac{g}{k}\tanh{(kh)}.
\end{equation}

It is not the case here, and we must use the air pressure evaluated at $z=\eta$ to determine $c$.
\subsection{The air domain}
\label{air domain}
Let us consider the linearized (inviscid) governing equation of a steady air 
flow, with a prescribed mean horizontal velocity $U(z)$ depending on the 
vertical coordinate $z$. We are going to study perturbations to the mean 
flow $U(z)$: $u_a(x,z,t)$, $w_a(x,z,t)$ and $P_a(x,z,t)$ 
(subscript $a$ stands for \emph{air}). So with 
$\mathbf{P}_a(x,z,t)=P_a(x,z,t) + \rho_a gz -P_{0} $, $\rho_a$ the air 
density, and $U'= dU(z)/dz$ we have the following equations
\begin{eqnarray}
u_{a,x}+w_{a,z}&=&0,\label{continuityair}\\
 \rho_a[u_{a,t} + U(z)u_{a,x} + U'(z)w_a]&=&-\mathbf{P}_{a,x},\label{uair}\\
\rho_a[w_{a,t} + U(z)w_{a,x}] &=&-\mathbf{P}_{a,z},\label{wair}
\end{eqnarray}
which must be completed with some appropriate boundary conditions. The first one is the kinematic boundary condition for air,
evaluated at the aerodynamic sea surface roughness $z_0$ placed just above the interface. 
It reads
\begin{equation}
 \eta_t + U(z_0)\eta_x = w_a(z_0) \label{etaair}.
 \end{equation}
We choose $U(z)$ to be the logarithmic wind profile. The log wind profile is 
commonly used to describe the vertical distribution of the horizontal mean 
wind speed within the lowest portion of the air-side of the marine boundary 
layer \citep{Garratt}

\begin{equation}\label{Udefiniton}
U(z) = U_{1} \ln(z/z_{0}),\quad U_{1} = \frac{u_{*}}{\kappa},\quad \kappa 
\approx 0.41,
\end{equation}
where $u_{*}$ is the friction velocity\footnote{Scaling the wind speed with the friction velocity is not a keystone of this work, and the reference speed can be changed.} and $\kappa$ the Von K\'{a}rm\'{a}n constant. So, eq. (\ref{etaair}) can be reduced to 
\begin{equation}\label{etaairreduced}
\eta_t = w_a(z_0).
\end{equation}

This equation describes the influence of the surface perturbation on the 
vertical perturbed wind speed. Next
we assume $\mathbf{P}_a=\mathcal{P}_a(z)\exp{(i\theta)}$, $u_a=\mathcal{U}_a(z)
\exp{(i\theta)}$,  $w_a=\mathcal{W}_a(z)\exp{(i\theta)}$ and we add the following 
boundary conditions on $\mathcal{W}_a$ and $\mathcal{P}_a $, 
\begin{align}
\lim\limits_{ z \to +\infty}(\mathcal{W}_a' + k\mathcal{W}_a) &= 0, \label{W_ainfinite}\\
\lim\limits_{ z \to z_0} \mathcal{W}_a &= W_{0},\label{W_ainz0}\\
\lim\limits_{ z \to+\infty} \mathcal{P}_a &= 0, \label{P_ainfinite}
\end{align}
that is, the disturbance vanishes at infinity, and the vertical component of 
the wind speed is enforced by the wave movement at the sea surface.
Then, using
equations (\ref{continuityair})-(\ref{wair}) and (\ref{P_ainfinite}) we obtain
\begin{align}
 w_a(x,z,t)&=\mathcal{W}_a\exp{(i\theta)},\label{w_a}\\
 u_a(x,z,t)&=\frac{i}{k}\mathcal{W}_{a,z}\exp{(i\theta)},\label{u_a}\\
 \mathbf{P}_a(x,z,t)&=ik\rho_a\exp{(i\theta)}\int^{\infty}_{z}[U(z')-c]
 \mathcal{W}_{a}(z') \text{d}z'.\label{mathP_a}
\end{align}
Removing the pressure from the Euler equations, we find the well-known Rayleigh equation (inviscid Orr-Sommerfeld)
\begin{equation}
 (U - c) (\mathcal{W}_a'' - k^2 \mathcal{W}_a) - U'' \mathcal{W}_a = 0 \quad \forall z~ \backslash~ z_{0} < z < 
+\infty,\label{Rayleigh}
\end{equation}
which is singular in $z_{c} = z_0 e^{c \kappa/ u_{*}} > z_{0} > 0$, where 
$U(z_{c}) = c$. In equations (\ref{w_a})-(\ref{Rayleigh}) neither $\mathcal{W}_a(z)$ nor $c$ 
are known. 
In order to find $c$, we have to calculate
$P_a(x,\eta,t)$. We obtain
\begin{equation}\label{integralforP}
P_a(x,\eta,t)=P_0 - \rho_a g \eta+ {ik\rho_a\exp{(i\theta)}\int^{\infty}_{z_0}[U(z)-c]\mathcal{W}_a(z)\text{d}z},
\end{equation}
where the lower integration bound is taken at the roughness height $z_0$ 
instead of $z=\eta$ since we are studying the linear problem. Finally, using
equation (\ref{etaairreduced}) to eliminate the term $ik\rho_a\exp(i\theta)$
the equation (\ref{integralforP}) in (\ref{deterc}) yields
\begin{equation}\label{equationforc}
g(1-s) + c\frac{sk^2}{W_0}\int^{\infty}_{z_0}U(z)\mathcal{W}_a(z)\text{d}z-c^2\{\frac{sk^2}{W_0}
\int^{\infty}_{z_0}\mathcal{W}_a(z)\text{d}z + k\coth(kh)\}=0,
\end{equation}
where $s=\rho_a/\rho_w$. This is the dispersion relation of the problem.
If $h \tti$ we obtain the expression (3.7) found by \citet{BejiN}.
The parameter $s$ is small ($\rho_a/\rho_w \sim 10^{-3}$) and
(\ref{equationforc}) may be approximated as
\begin{equation}
c=c_0 + sc_1 + O(s^2).
\end{equation}
The explicit expression of $c_1$ is calculated in the next section. 
Therefore, we can find $\mathcal{W}_a(z)$ by solving (\ref{Rayleigh}) with $c$ substituted by $c_0$, that is to say, of order zero in $s$. The method is shortly described in 
appendix \ref{Orr}.
\section{Wave growth rate}
\label{growth}
The function $\mathcal{W}_a(z)$  
is complex and consequently $c$ also. Its imaginary part gives the growth rate of $\eta(x,t)$ defined by
\begin{equation}
 \gamma =k\Imag{(c)},
\end{equation}
where $\Imag{(c)}$ is the imaginary part of $c$.
The theoretical and numerical results concerning the growth rate $\gamma$ are studied and computed in terms of two dimensionless parameters $\delta$ (see \citet{YoungVerhagen1} and \citet{YoungVerhagen2}) and $\theta_{dw}$ defined by
\begin{equation}\label{delta}
\delta = \frac{gh}{U^2_1},\quad \theta_{dw}=\frac{1}{U_1}\sqrt{\frac{g}{k}}.
\end{equation} 
The dimensionless parameter $\delta$ is the more important novelty in relation to the set of non-dimensional parameters governing $\gamma$ in deep water.  This parameter measures the influence of the finite fluid depth on the rate of growth of
 $\eta(x,t)$. The parameter $\theta_{dw}$ is nothing more than the deep water wave age.
Now a {\it finite depth wave age} $\theta_{fd}$ can be introduced as
\begin{equation}\label{waveagefinite}
\theta_{fd}=\frac{1}{U_1}\sqrt{\frac{g}{k}}\sqrt{\tanh({kh})}=
\theta_{dw} T^{1/2},
\end{equation}
where $T = \text{tanh}(\frac{\delta}{\theta_{dw}^2})$. The form (\ref{waveagefinite}) defines a {\it depth weighted wave age} such that: for a finite and constant
$\theta_{dw}$ we have $\theta_{fd}\sim \theta_{dw}$ if $\delta \tti$ and $\theta_{fd}\sim \delta^{1/2}=
\sqrt{gh}/U_1$ if $\delta \rightarrow 0$. In order to obtain the growth rate, we introduce
the following non-dimensional variables and scalings (hats mean dimensionless quantities)
\begin{eqnarray}\label{parameters}
 U&=&U_1\hat{U},\quad \mathcal{W}_a=W_0\mathcal{\hat{W}}_a,
\quad z=\frac{\hat{z}}{k}, \quad c = U_{1} \hat{c},\quad t = \frac{U_{1}}{g} \hat{t}.
\end{eqnarray}
Using (\ref{delta}) and (\ref{parameters}) in equation (\ref{equationforc}) and retaining only the terms of order one in $s$ we obtain $c$,
 \begin{equation}
 \hat{c}=\hat{c}(\delta,\theta_{dw})=\theta_{dw}T^{1/2} -\frac{s}{2} \theta_{dw}T^{1/2} +\frac{s}{2}\{ T\hat{I}_1 -
 \theta_{dw} T^{3/2} \hat{I}_2\},
 \end{equation}
and with $e^{\gamma t} = e^{k {\Imag}(c) t} = e^{\Imag{(\hat{c})} \hat{t}/\theta_{dw}^{2}}$, we have the dimensionless growth rate $\hat{\gamma} = \frac{U_{1}}{g}\gamma$ as (dropping the hats)
\begin{equation}
 \gamma = \frac{s}{2}\{ \frac{T \Imag{(I_1)}}{\theta^2_{dw}}-
 \frac{T^{3/2}\Imag{(I_2)}}{\theta_{dw}}\},
\label{growthrate}
\end{equation}
where the integrals are defined as follow
\begin{equation}
I_1=\int_{z_0}^{\infty} U\mathcal{W}_a \text{d}z,\quad I_2=\int_{z_0}^{\infty}\mathcal{W}_a 
\text{d}z.
\end{equation}
So, we can compute it for a given $(\delta,\theta_{dw})$ set.
The $\delta$ parameter does not appear explicitly 
allowing us to indeed compute $\gamma$ for an infinite depth, where we have just $T\rightarrow 1$. This gives back Miles' theory.  
\section[Results]{Discussion}
\label{Results}
\begin{figure}
\begin{center}
 \includegraphics{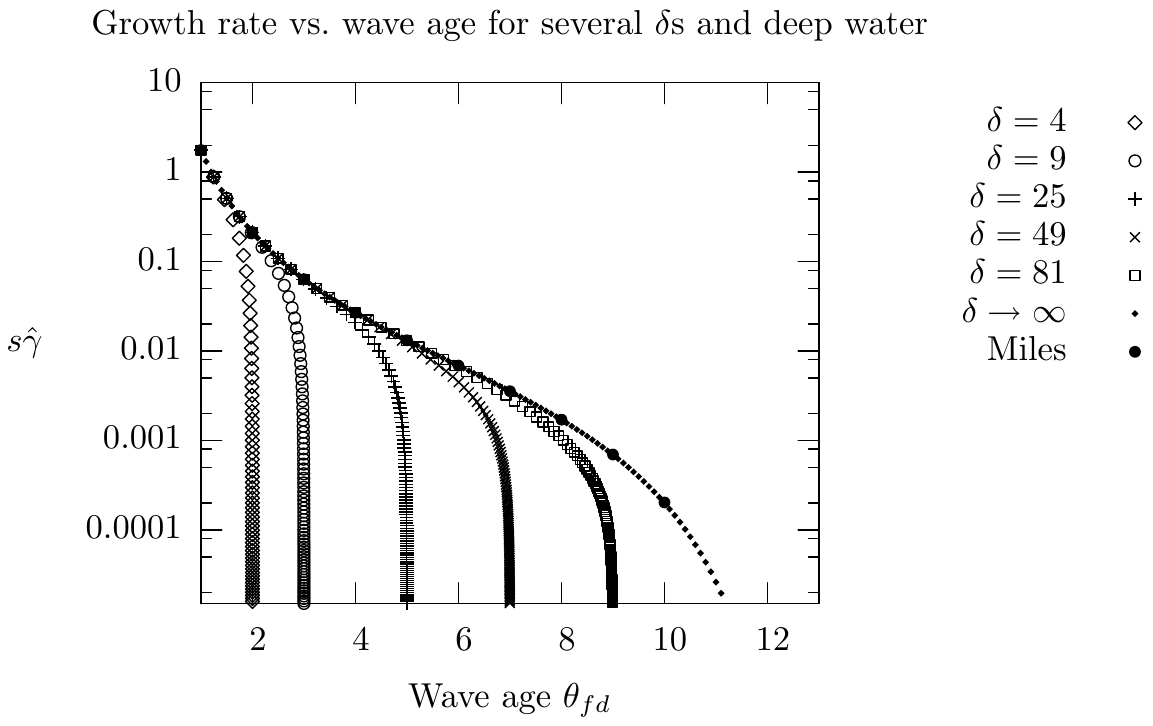}
\end{center}

\caption{Evolution of the growth rate in semi-logarithmic scale. The curves that goes to zero almost straightly are the finite depth ones. From left to right, they match $\delta =4,9,25,49,81$. We see that for each depth, there is a wave-age 
limited growth. The true deep water limit, also computed, is approached for 
small $\theta_{fd}$ and matches Miles' results.}
\label{fig_growth_rate}
\end{figure}

The existence of a finite depth $h$
transforms the unique curve of wave growth rate in deep water in a \textit{family of
curves} indexed by $\delta = gh/U^2_1$, i.e., a curve for each value of $\delta$.
In Figure \ref{fig_growth_rate} is shown a family of five values of $\delta$ 
against the finite-depth wave age. The limit $ \delta \rightarrow \infty$,  corresponding to the 
envelope of the family of curves, is included as well.

Figure \ref{fig_growth_rate} clearly shows that at small wave age the growth
rate $\gamma$ is almost equal for all values of $\delta$, the limit being the deep water case.

As the wave age increases, the finite-depth effects begin to appear. The growth rate becomes lower than in the deep water limit, for each value of $\delta$. The growth rates are scaled with $\delta$: for a given wave age, the bigger the $\delta$ the larger the $\gamma$. Each $\delta$-curve approaches its own (idealized) \emph{finite-depth wave-age-limited growth} as $\gamma$
goes to zero. At this stage the wave reaches a final state of linear progressive wave with zero 
growth. In others words, for a given $\delta$ the surface wave does not grow 
old anymore beyond a determined wave age.

We note that the theoretical curves in Figure \ref{fig_growth_rate} are,  
$\textit{mutatis mutandi}$, in good agreement with the empirical 
curves of the fractional wave energy increase per radian $\Gamma$ as a function
of the inverse wave age $U_{10}/Cp$
in \citet{Young1} ( see Figs. $3a$, $3b$ $3c$ and $3d$).

\begin{figure}
\begin{center}
\includegraphics{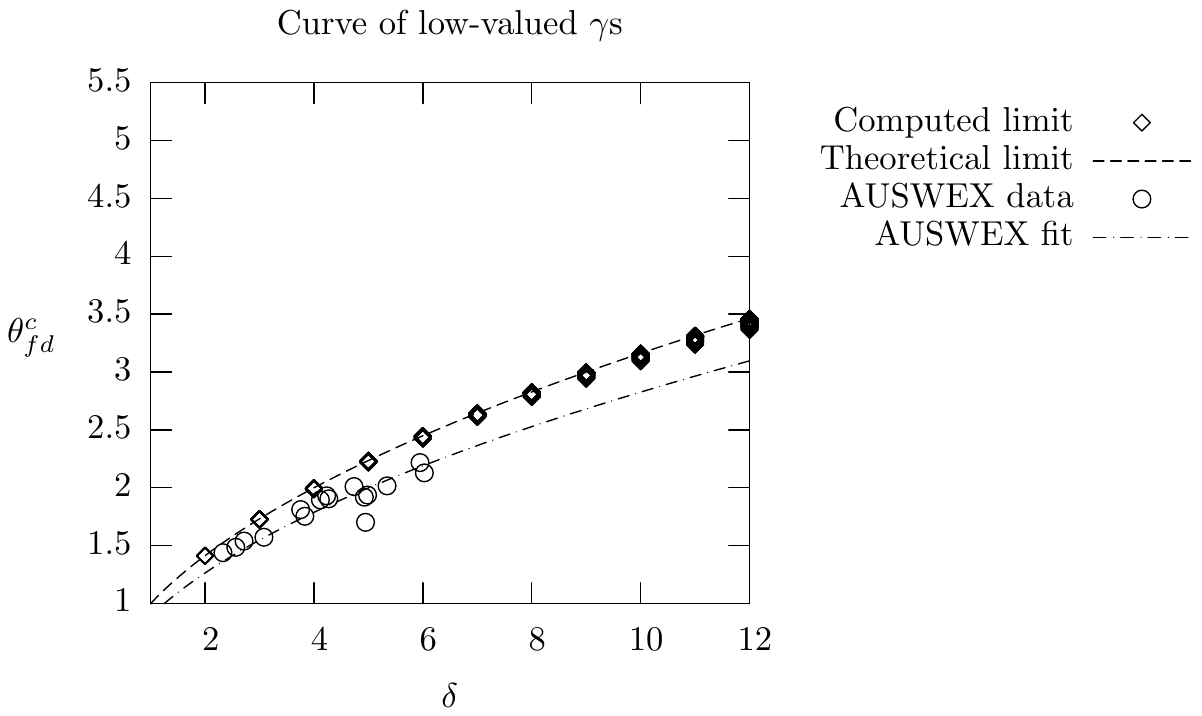}
\end{center}
\caption{Parameter curves where the growth rate is almost zero. The theoretical limit is (\ref{thetacritique}). The AUSWEX data was taken from \citet{Donelan}, and the AUSWEX fit is $\theta_{fd} = 0.8 \delta^{1/2}$.}
\label{gamma_zero} 
\end{figure}

In Figure \ref{gamma_zero} are plotted, against $\delta$, the critical values of the wave age $\theta_{fd}^{c}$ for which the growth rate $\gamma$ goes to zero. They obey the relation
\begin{equation}
\label{thetacritique}
 \theta_{fd}^{c} = \delta^{1/2}.
\end{equation}
The above relation, found numerically, is coherent with the parameter formulation (\ref{waveagefinite}). It is indeed a limiting value for the wave age, uniquely determined by the water depth.
This result is in excellent agreement with the one given by \citet{Young1} for $\Gamma$ (from \citet{Donelan2}) The authors have shown from an empirical relationship (formule (6) in reference above) that $\Gamma$ goes to zero for
\begin{equation}
 \frac{C_p}{U_{10}}=0.8(\frac{gh}{U_{10}^{2}})^{0.45}.
\end{equation}
In Figure  \ref{gamma_zero} are also represented data from \citet{Donelan}, from the AUstralian Shallow Water EXperiment, wich will be referred to as AUSWEX. A basic fit is also plotted to show the general trend. The raw data consists in the water depth $h$ in metres, the friction velocity $u_{*}$, the $10$ metres wind velocity $U_{10}$ and the ratio of the former with the measured phase speed $c_p$, $U_{10}/c_p$. For example, $u_{*} = 0.44~\text{m}.\text{s}^{-1}$, $h = 0.32~\text{m}$, $U_{10} = 11.9~\text{m}.\text{s}^{-1}$ and $U_{10}/c_p = 7.2$ gives $\delta = 2.7$ and $\theta_{fd} = 1.55$, which gives a small relative error regarding (\ref{thetacritique}). All the points give $(\delta,\theta_{fd})$ coordinates really close to the theoretical limit.

With $\theta_{fd}^{c}$ we can calculate the corresponding critical wave length $\lambda^{c}$. Using 
(\ref{thetacritique}) in (\ref{waveagefinite}) we obtain
\begin{equation}\label{shallow}
 \frac{\delta}{\theta^2_{dw}}=\tanh{(\frac{\delta}{\theta^2_{dw}})}.
\end{equation}
Relation (\ref{shallow}) means the wave has entered the shallow water region.
In such a limit the range of  $\delta/\theta^2_{dw}$ is: $0<\delta/\theta^2_{dw}<\frac{\pi}{4}$ (\citet{Fenton}, \citet{Marcus}). As a result we obtain $\lambda > \lambda^{c} = 8 h$.
For these values of $\lambda$, the phase velocity is in the long wave limit i.e., $c=\sqrt{g h}$. Consequently, if $\lambda > \lambda_c$ the wave feels the bottom, the amplitude does not grow anymore, the resonance wind/phase speed ceases, and the wave reaches its utmost state as a progressive plane wave.

An analogous phenomenon appears in the deep water limit $\delta \rightarrow \infty$ where $\tanh{(\frac{\delta}{\theta^2_{dw}})}$ is in $[0.99;1]$
for $\delta/\theta^2_{dw} > \pi$, so we have
\begin{equation}\label{gammazerodeep}
\gamma=\frac{s}{2}\{ \frac{\Imag{(I_1)}}{\theta^2_{dw}}-
\frac{\Imag{(I_2)}}{\theta_{dw}}\}.
\end{equation}
The $\lambda^{c}$ is in this case $\lambda^{c}=2h$. For $\lambda <\lambda^{c} $,  
 $\gamma$ goes to $0$ and the sea reaches the final state for which we can obtain from (\ref{gammazerodeep}) the phase velocity $\theta_{dw}$
\begin{equation}
 \theta_{dw}= \frac{1}{U_1}\sqrt{\frac{g\lambda}{2\pi}}=\frac{\Imag{(I_1)}}{\Imag{(I_2)}}=\frac{\Imag{(\int^{\infty}_{z_0}U(z)\mathcal{W}_a(z)\text{d}z})}
{\Imag(\int^{\infty}_{z_0}\mathcal{W}_a(z)\text{d}z)}.
\label{limiting_theta}
\end{equation}

\begin{figure}
\begin{center}
\includegraphics{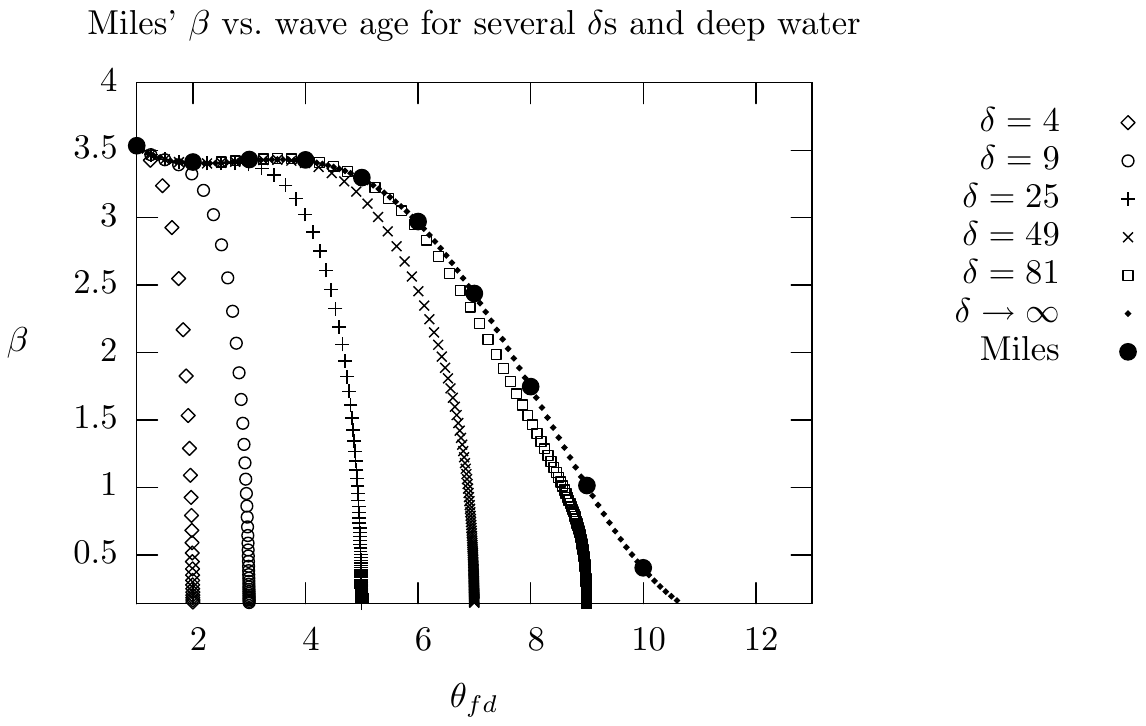}
\end{center}
\caption{Evolution of Miles' $\beta$. Every curve is calculated with the same $z_0$, the differences are only due to the various depths.  As we see, the finite-depth effect is critical, and for high $\delta$, we reach a deep water behaviour.}
\label{fig_growth_rate_beta}
\end{figure}

In contrast to the usual analysis in wind-induced wave growth (in deep or finite
depth), our results concern the dimensionless growth rate 
$\gamma$ in
terms of the finite depth wave age $\theta_{fd}$ instead of a finite depth fetch 
$\chi_{fd}$. Therefore a transformation rule is needed. It can be easily
derived from the empirical relation introduced by Hasselmann \citep{Hasselman}
\begin{equation}
\frac{f_pU_{10}}{g}=3.5\left(\frac{xg}{U^2_{10}}\right)^{D},\quad D \approx -0.33,
\end{equation}
with $f_p$ the peak frequency, $U_{10}$ the wind speed at $10~m$ and $x$ the
fetch in metres. Now using $f_p=g/(2\pi c_0)$, $\delta = \frac{gh}{U^2_1}$ and the
formula in (\ref{Udefiniton}) we obtain a transformation between 
$\theta_{fd}$ and the dimensionless fetch $\chi_{fd}$
\begin{equation}
\theta_{fd} =\frac{1}{7\pi} \left(\ln\left(\ \frac{gz_{10}}
{\alpha_C U^2_1\kappa^2} \right)\right)^{1+2D}(\chi_{fd})^{-D}, \quad \chi_{fd}=\frac{xg}{U^2_1},
\end{equation}
with $z_{10}=10~m$ and the Charnock constant $\alpha_C = 0.018$ 
used to determine the roughness length $z_0=\alpha_C u^2_{*}/g$. This relation was proposed by \citet{Charnock}, and is used since, although other parameterizations of the roughness length exist. Johnson's relation, for example, states a dependence between $z_0$ and the wave age \citep{Johnson}.
Identically we have the following transformation rule between the Miles' dimensionless growth rate $\beta$ and dimensionless $\gamma$  
\begin{equation}
\beta=\frac{2\gamma}{s}\theta_{dw}^{3} T^{1/2},
\end{equation}
where we took $\beta$ as it is usually defined, with the dimensions,
$\Imag(c) = c_{0} \frac{s}{2} \beta (\frac{U_{1}}{c_{0}})^{2}.$
This is a straightforward definition of Miles' $\beta$ in finite depth. Its evolution is shown clearly in Figure \ref{fig_growth_rate_beta}, showing the usual deep water trends, and the new finite depth limits. The effects of depth are critical. The $\beta$ is almost constant for small $\theta_{fd}$, as usual, but it goes to zero extremely fast when the finite-depth wave-age limit is close.

\section{Conclusions and Perspectives}
\label{conclusions}
In this work we built a theoretical extension to the finite depth domain of Miles' theory of wave generation by wind (\citet{miles2}). In order to compute wave growth rates in finite depth we adapted the method introduced by  \citet{BejiN} for deep water. We defined a finite depth wave age
useful to study families of wave growth rates, each member of the family tagged by a water depth.

From a numerical study based on purely theoretical relationships it is shown that wave growth rates go to zero for long waves in shallow water.
 For small wave age the wave growth rates behave as in deep water, regardless of the actual depth.
These results are the qualitative analogous of 
the well known empirical (or semi-empirical) results concerning wave energy increase per radian in function of the inverse wave age (\citet{Young1},\citet{Young2} ). We have shown, for the first time, curves of the original Miles' dimensionless growth rate $\beta$ in finite depth. This growth rate drops from a deep water behaviour to zero extremely quickly as the wave age reaches its limit.

Miles' theory studies exponential growth of the wave amplitude of a linear and uniform monochromatic wave train.
Nowadays nonlinear modulational analysis beyond Miles' 
monochromatic theory were carried out. 
In reference \citet{Thomas}, the authors report the behaviour of the  
Benjamin Feir
instability when dissipation and wind input are both taken into account. 
Within the framework of weakly nonlinear modulated wave train, with this same theory as the foundations, we are going to study the influence of the wind in the
finite depth Benjamin-Feir instability (for $kh>1.363$). Inclusion of small bottom friction and constant vorticity are also under consideration.

~\\~\\~\\
{ \paragraph{}Acknowledgements  P. M. thanks  \textbf{Labex NUMEV} (\textit{Digital and Hardware Solutions, Modelling
for the Environment and Life Sciences}) for partial financial support.}

\newpage
\appendix
\section{Rayleigh equation}
\label{Orr}
We recall the Rayleigh equation
\begin{equation}
(U - c) (\mathcal{W}_a'' - k^2 \mathcal{W}_a) - U'' \mathcal{W}_a = 0 \quad \forall z~ \backslash~ z_{0} < z < +\infty
\label{Rayleighnum}
\end{equation}
which is singular in $z_{c} > z_{0} > 0$, where $U(z_{c}) = c$. This equation underlies the most essential mechanism of flow stability. This equation is singular only for the zeroth order in $s$, where $c = c_0 + o(s)$, which is real. The value of $c_1$, hopefully complex, is then found with the dispersion relation (\ref{equationforc}). Nevertheless, this is computational, and do not change the fact that we search a complex $c$ eigenvalue in (\ref{Rayleighnum}).
We have to prescribe a flow $U(z)$ allowing instability, or more precisely, not forbidding it. Rayleigh's inflexion point theorem, and subsequent Fjortoft theorem states that $U(z)$ is bound to have one inflexion point to, at least, not forbid instability (see \citet{Fjortoft}). Usually, for this theorem, the domain of $\mathcal{W}_a$ is $[0,+\infty[$, and the boundary conditions are the vanishing at each bound. But our lower bound is $z_0$, which is nonzero, and we have a forcing of $\mathcal{W}_a$ in $z_0$. Taking this into account, we derive the following constraint
\begin{align}
 \Imag{(c)} \int_{z_0}^{+\infty} \frac{U''(z)}{|U(z) - c|^2} |\mathcal{W}_a(z)|^2 \text{d}z = -\Imag(\lim\limits_{ z \to z_0}\mathcal{W}_a^{*}(z)\mathcal{W}_a'(z)).
\end{align}
Where we see that if $z_0 \ttz$ and $\mathcal{W}_a$ vanishes smoothly at the boundaries, considering that $U''(z)$ is monotonous, indeed the r.h.s vanishes and $\Imag(c)$ must be zero. Here, it can be nonzero.
So, the condition found allows an exponential growth of the free surface $\eta(x,t)$, like a mechanical oscillating system forced into one of its normal modes \citep{ConteMiles}. 
We use now a semi-numerical recipe to solve (\ref{Rayleighnum}) for $\mathcal{W}_a$ following the method introduced by
Beji and Nadaoka. 
We first develop (\ref{Rayleighnum}) in the $z_{c}$-neighbourhood, assuming
\begin{align}
 U(z) &\approx U'(z_{c}) ( z- z_{c}) + c, \quad U''(z) \approx U''(z_{c}),\quad
 (k \frac{U'(z_{c})}{U''(z_{c})})^{2}\rightarrow 0.
\label{Vital_Assumptions}
\end{align}
These are fairly true in the logarithmic wind profile case. After some algebra,
 these assumptions transform (\ref{Rayleighnum}) in a Bessel equation of order 
 one, whose solutions are known to be a linear combination of the two 
 first-order Bessel functions. So, with the weighted-centered, dimensionless 
 variable $z_{p} = -U''(z_{c})\frac{(z - z_{c})}{U'(z_{c})} = \frac{z}{z_c} - 1$
 we find
\begin{equation}
 \mathcal{W}_{a}(z_{p}) = \sqrt{z_{p}} \left( C_{1} J_{1}(2 \sqrt{z_{p}}) + C_{2} Y_{1}(2\sqrt{z_{p}} )\right).
\label{Solution_Bessel}
\end{equation}
Where $C_{1}$ and $C_{2}$ are complex constants. Here, the function $\mathcal{W}_a$ is $z_p$-dependent. The numerical solution that we seek can be written as 
\begin{equation}
 \mathcal{W}_a = C_1 W_{1} + C_2 W_{2},
\label{Sol_num}
\end{equation}
where $W_{1}$ and $W_{2}$ are unknown and independent. $W_{1}$ (resp. $W_{2}$) is found by integrating (\ref{Rayleighnum}) with $J_{1}$ (resp. $Y_1$) for initial function value and slope value at $z_c$.
The first thing we need is to avoid the numerical singularity. We achieve this by introducing a small parameter $\varepsilon$. We choose the two initial points to be $z_{c}^{\pm} = (1\pm\varepsilon) z_{c}$, and evaluate the Bessel functions in these points.
As we notice, the function $z_p^{1/2} Y_{1} (2 z_p^{1/2})$ becomes complex, with a negative imaginary part, when $z<z_c$. Its derivative becomes complex, with a positive imaginary part.

 We have to take the complex-conjugates of the initial value and initial slope value of these functions to get a positive growth rate\footnote{Taking the non-conjugate values will simply give a minus sign to $\Imag(c)$.} \citep{DrazinReid}. Therefore, after integration, we get $W_1$ and $W_2$.
But this doesn't take into account integration constants. These are set using the boundary conditions of surface forcing and vanishing at infinity. So, using (\ref{Sol_num}) in (\ref{W_ainfinite}) and (\ref{W_ainz0}), we obtain a simple algebraic system
which allows us to determine $C_1$ and $C_2$. The ``infinite'' value, for computation, is the one from which the constants are stable enough. The relative error on the constants is proportional to the error on the growth rate. Here, a relative error of $10^{-5}$ is taken.
As a result, we obtain $\mathcal{W}_a(z)$, and we can evaluate the integrals
 in (\ref{growthrate}) for any parameters set. The results are shown in 
Figure \ref{fig_growth_rate}.

\newpage
\nocite{*}
\bibliographystyle{jfm}
\bibliography{./biblio}

\begin{thebibliography}{27}
\expandafter\ifx\csname natexlab\endcsname\relax\def\natexlab#1{#1}\fi

\bibitem[Beji \& Nadaoka(2004)]{BejiN}
{\sc Beji, S. \& Nadaoka, K.} 2004 Solution of rayleigh's instability equation
  for arbitrary wind profiles. {\em J. Fluid Mech.\/} {\bf 500}, 65--73.

\bibitem[Charnock(1955)]{Charnock}
{\sc Charnock, H.} 1955 Wind stress on a water surface. {\em Quart. J. Roy.
  Meteorol. Soc.\/} {\bf 81}, 639--640.

\bibitem[Conte \& Miles(1959)]{ConteMiles}
{\sc Conte, S.~D. \& Miles, J.~W.} 1959 On the numerical integration of the
  orr-sommerfeld equation. {\em Journal of Society of industrial and applied
  mathematics\/} {\bf 7}~(4), 361--366.

\bibitem[Donelan {\em et~al.\/}(2006)Donelan, Babanin, Young \&
  Banner]{Donelan}
{\sc Donelan, M.~A., Babanin, A.~V., Young, I.~R. \& Banner, M.~L.} 2006
  Wave-follower field measurements of the wind-input spectral function. part ii
  : Parameterization of the wind input. {\em Journal of Physical
  Oceanography\/} {\bf 36}, 1672--1689.

\bibitem[Donelan {\em et~al.\/}(1992)Donelan, Skafel, Graber, Liu, Schwab \&
  Venkatesh]{Donelan2}
{\sc Donelan, M.~A., Skafel, M., Graber, H., Liu, P., Schwab, D. \& Venkatesh,
  S.} 1992 On the growth rate of wind-generating waves. {\em Atmos.-Ocean\/}
  {\bf 30}, 457--478.

\bibitem[Drazin \& Reid(1982)]{DrazinReid}
{\sc Drazin, P.~G. \& Reid, W.} 1982 {\em Hydrodynamic Stability\/}. Cambridge
  University Press.

\bibitem[Fenton(1979)]{Fenton}
{\sc Fenton, J.~D.} 1979 A high-order cnoidal wave theory. {\em J. Fluid
  Mech.\/} {\bf 94}, 129--161.

\bibitem[Fjortoft(1950)]{Fjortoft}
{\sc Fjortoft, R.} 1950 Application of integral theorems in deriving criteria
  of stability of laminar flow and for the baroclinic circular vortex. {\em
  Geofys. Pub.\/} {\bf 17}, 1--52.

\bibitem[Francius \& Kharif(2006)]{Marcus}
{\sc Francius, M. \& Kharif, C.} 2006 Three-dimensional instabilities of
  periodic gravity waves in shallow water. {\em J. Fluid Mech.\/} {\bf 561},
  417--437.

\bibitem[Garratt {\em et~al.\/}(1996)Garratt, Hess, Physick \&
  Bougeault]{Garratt}
{\sc Garratt, J.~R., Hess, G.~D., Physick, W.~L. \& Bougeault, P.} 1996 The
  atmospheric boundary layer advances in knowledge and application. {\em
  Boundary-Layer Meteorology\/} {\bf 78}, 9--37.

\bibitem[Hasselmann {\em et~al.\/}(1973)Hasselmann, Barnett, Bouws, Carlson,
  Cartwright, Eake, Euring, Gicnapp, Hasselmann, Kruseman, Meerburg,
  {M{\"{u}}ller}, Olbers, K.Richter, Sell \& Walden]{Hasselman}
{\sc Hasselmann, K., Barnett, T.~P., Bouws, E., Carlson, H., Cartwright, D.~E.,
  Eake, K., Euring, J.~A., Gicnapp, A., Hasselmann, D.~E., Kruseman, P.,
  Meerburg, A., {M{\"{u}}ller}, P., Olbers, D.J., K.Richter, Sell, W. \&
  Walden, H.} 1973 Measurements of wind-wave growth and swell decay during the
  joint north sea wave project (jonswap). {\em Ergnzungsheft zur Deutschen
  Hydrographischen Zeitschrift Reihe\/} {\bf 8}~(12), 95.

\bibitem[Janssen(2004)]{Janssen}
{\sc Janssen, P.A.E.M} 2004 {\em The interaction of ocean waves and wind\/}.
  Cambridge University Press.

\bibitem[Jeffreys(1924)]{Jeffreys1}
{\sc Jeffreys, H.} 1924 On the formation of waves by wind. {\em Proc. Roy.
  Soc.\/} {\bf A107}, 189--206.

\bibitem[Jeffreys(1925)]{Jeffreys2}
{\sc Jeffreys, H.} 1925 On the formation of waves by wind. ii. {\em Proc. Roy.
  Soc.\/} {\bf A110}, 341--347.

\bibitem[Johnson {\em et~al.\/}(1998)Johnson, H{\o}jstrup, Vested \&
  Larsen]{Johnson}
{\sc Johnson, H.~K., H{\o}jstrup, J., Vested, H.~J. \& Larsen, S.E.} 1998 On
  the dependence of sea surface roughness on wind waves. {\em J. Phys.
  Ocean.\/} {\bf 28}, 1702--1716.

\bibitem[Kharif {\em et~al.\/}(2010)Kharif, Kraenkel, Manna \& Thomas]{Thomas}
{\sc Kharif, C., Kraenkel, R., Manna, M.~A. \& Thomas, R.} 2010 The modulation
  instability in deep water under the action of wind and dissipation. {\em J.
  Fluid Mech.\/} {\bf 664}, 138--149.

\bibitem[Lighthill(1978)]{Lighthill}
{\sc Lighthill, M.~J.} 1978 {\em Waves in Fluids\/}. Cambridge University
  Press.

\bibitem[Miles(1957)]{miles2}
{\sc Miles, J.~W.} 1957 On the generation of surface waves by shear ﬂows.
  {\em J. Fluid Mech.\/} {\bf 3}, 185--204.

\bibitem[Miles(1997)]{Miles3}
{\sc Miles, J.~W.} 1997 Generation of surface waves by winds. {\em Appl. Mech.
  Rev\/} {\bf 50-7}, R5--R9.

\bibitem[Phillips(1957)]{Phillips}
{\sc Phillips, O.~M.} 1957 On the generation of waves by turbulent wind. {\em
  J. Fluid Mech.\/} {\bf 2}, 417--445.

\bibitem[Pierson~Jr \& Moskowitz(1964)]{PiersonMo}
{\sc Pierson~Jr, W.~J. \& Moskowitz, L.} 1964 A proposed spectral form for
  fully developed wind seas based on the similarity theory of s.a.
  kitaigorodskii. {\em Journal of geophysical research\/} {\bf 69}~(24),
  5181--5189.

\bibitem[Rayleigh(1880)]{Rayleigh}
{\sc Rayleigh, Lord} 1880 On the stability, or instability, of certain fluid
  motions. {\em Proc. Lond. Math. Soc\/} {\bf XI}, 57--70.

\bibitem[Touboul {\em et~al.\/}(2008)Touboul, Kharif, Pelinovsky \&
  Giovanangeli]{TouboulKharif1}
{\sc Touboul, J., Kharif, C., Pelinovsky, E. \& Giovanangeli, J-P} 2008 On the
  interaction of wind and steep gravity wave groups using miles' and jeffreys'
  mechanisms. {\em Nonlinear Processes in Geophysics\/} {\bf 15}~(6),
  1023--1031.

\bibitem[Young(1997)]{Young1}
{\sc Young, I.~R.} 1997 The growth rate of finite depth wind-generated waves.
  {\em Coastal Eng.\/} {\bf 32}, 181--195.

\bibitem[Young(1999)]{Young2}
{\sc Young, I.~R.} 1999 {\em Wind Generated Ocean Waves\/}. Elseiver.

\bibitem[Young \& Verhagen(1996a)]{YoungVerhagen1}
{\sc Young, I.~R. \& Verhagen, L.~A.} 1996a The growth of fetch limited waves
  in water of finite depth. part i : Total energy and peak frequency. {\em
  Coastal Eng.\/} {\bf 29}, 47--78.

\bibitem[Young \& Verhagen(1996b)]{YoungVerhagen2}
{\sc Young, I.~R. \& Verhagen, L.~A.} 1996b The growth of fetch limited waves
  in water of finite depth. part ii : Spectral evolution. {\em Coastal Eng.\/}
  {\bf 29}, 79--99.

\end{thebibliography}
\end{document}